\documentstyle[mprocl]{article}

\def\be{\begin{equation}}
\def\ee{\end{equation}}
\def\bea{\begin{eqnarray}}
\def\eea{\end{eqnarray}}

\begin{document}

\title{D-Particles, D-Instantons, and A  Space-Time Uncertainty 
Principle   \\in String Theory}

\author{ Tamiaki YONEYA }

\address{Institute of Physics\\Graduate School of 
Arts and Sciences\\
University of Tokyo,
Komaba, Meguro-ku, Tokyo 153\\
E-mail: tam@hep1.c.u-tokyo.ac.jp}   




\newcommand{\EQ}{\begin{equation}}
\newcommand{\EN}{\end{equation}}
\newcommand{\EQA}{\begin{eqnarray}}
\newcommand{\EQN}{\end{eqnarray}}
\newcommand{\e}{{\rm e}}
\newcommand{\Sp}{{\rm Sp}}
\newcommand{\Tr}{{\rm Tr}}
\newcommand{\p}{\partial}

\maketitle\abstracts{
The purpose of this talk is to review 
 some considerations by the present author 
on the possible role of a simple space-time 
uncertainty relation toward  
nonperturbative string theory. 
We first motivate the space-time uncertainty 
relation as a simple space-time characterization of the 
fundamental string theory. 
In the perturbative string theory, the relation can be 
regarded as a direct consequence of the 
world-sheet conformal 
invariance. We argue that the relation 
captures some of the important 
aspects of the short-distance dynamics of D-particles 
described by the effective super Yang-Mills matrix quantum 
mechanics, and also that the recently proposed type IIB matrix model   
can be regarded  as the simplest realization of the 
space-time uncertainty principle.  
 }

\section{Introduction}
\subsection{Motivation: Basic principles 
behind string theory?}
Many people working in string theory now 
believe that the string theory as the prime candidate 
toward the unified quantum theory of all 
interactions including gravity has now reached 
at a very crucial stage. 
Namely, the recent developments related to 
the string dualities  completely changed 
the naive view on the theory: The theory 
is now not even a theory of strings, but rather it 
suggests some more fundamental degrees of freedom 
behind strings. The strings turn out to be  only one form 
of the various possible degrees of freedom of the theory 
which appears in the special perturbative 
regime of the theory. In fact, varieties of extended 
objects (called ``branes") with various dimensions 
appear in different regimes 
of the theory, and they may play equally important 
roles as the 
one-dimensional objects, strings. 
Although a  suspicion that the string may not be 
the fundamental objects of the theory has long been 
held by many people, this is the first time that we have 
concrete evidence and handles for that. We may now  
attack  the problem of formulating the string 
theory nonperturbatively in a well-defined manner.  

In spite of these impressive developments, 
however, we do not yet have any clear 
understanding on the basic principles 
which underlie the string theory.  
In view of this, it seems worthwhile to 
present some considerations on a new 
uncertainty principle with respect to 
space and time \cite{yoneya1}$^,$ \cite{yoneya2},
 which, albeit speculative and hence 
rather vague yet, 
explains in a very simple way some of the 
crucial qualitative aspects of both the 
perturbative and nonperturbative 
excitations of the theory and might play some 
role in the quest of basic principles of 
string theory.  

\subsection{Origin of the space-time uncertainty relation}

To motivate our discussion below, let us first 
briefly recall the meaning of the fundamental 
string scale. Obviously, one of the most significant 
feature of the string theory is that it introduces 
the fundamental unit of length, $\sqrt{\alpha'}
\equiv \ell_s$. The string length $\ell_s$ 
replaces the role of the Planck length in 
ordinary approaches to quantizing gravity 
within the framework of  local field theory. 
In string theory, 
the gravitational interaction  is built-in in a completely 
unified way with all other particle interactions. 
In the unit in which the  $10$ dimensional Newton constant and the 
light velocity 
are unity, the Planck constant $h$ is related to the string 
length as 
\EQ
h = g_s^2 \ell_s^8
\EN
  where $g_s$ 
is the dimensionless string coupling given 
by the vacuum expectation value of the dilaton 
$g_s = e^{\phi}$. Note that, in this ``gravity unit", 
the dimension of mass is $L^{D-3}$ in terms of the 
length dimension $L$, while the dimension of the Planck 
constant $\hbar$ is $L^{D-2}$, in $D$ dimensional space-time. 
Thus the classical limit $h \rightarrow 0$ keeping 
gravity finite corresponds either to $\ell_s \rightarrow 0$ or 
to $g_s\rightarrow 0$. The latter situation means that 
the universe to be described by the classical 
physics is composed of classically extended string objects. 
This does not fit our experience, and we are forced to 
adopt the former option. In this case, the introduction of the 
Planck's quantum is  
directly equivalent to introducing the string scale.      
From this viewpoint, the string theory should perhaps be 
interpreted as a theory of the quantized space-time, 
in which  the ordinary space-time continuum plays a somewhat analogous role 
as the classical phase space in ordinary quantum mechanics. 

Let us now try to reinterpret the string theory 
from this viewpoint. Since 
in any theory of quantum gravity 
we expect the appearance of a limiting length 
beyond which the ordinary concept of the 
continuum space-time should be invalidated, 
it is important to identify the manner how such 
a limitation arises in the fundamental string theory. 
One such characterization of the theory comes from 
a reinterpretation of the ordinary time-energy uncertainty 
relation. 
\EQ
\Delta T \Delta E \ge \hbar. 
\EN
Now the energy of a string is roughly proportional to 
the length of the string $X_l$, the string tension 
$T\sim {\hbar \over \alpha'}$ being the 
proportional constant:
\EQ
\Delta E \sim  {\hbar \over \alpha'}\Delta X_l .
\EN
Thus we have the relation for the 
uncertainties with respect to the time interval and the 
spatial length \cite{yoneya1} $^,$ \cite{yoneya2}, 
\EQ
\Delta T\Delta X_l \ge \ell_s^2 .
\label{stu}
\EN
It is tempting to assume this relation as a 
fundamental uncertainty relation which governs 
the short-distance behavior of the string theory as 
a ``theory of quantized space-time", since it expresses the 
limitation of the ordinary concept of the continuum space-time 
directly in terms of the string scale $\ell_s$ and the space-time area.
\footnote{Historical remark: 
This viewpoint was first suggested by the 
present author \cite{yoneya1} just a decade ago, but unfortunately 
has not been developed further at that time 
except for a later discussion given in \cite{yoneya2}. Independently of 
this suggestion, an intimately related concept of the `minimal length' 
was suggested by others \cite{ven}\cite{gross} and has got  
popularity.} 
The dual role of the time and the (longitudinal
\footnote{
Here it is important to discriminate the length scales in the longitudinal 
and transverse directions with respect to the string. 
The length scale appeared in the uncertainty relation 
(\ref{stu}) is the one along the longitudinal direction. 
 As is well known, the (squared) 
transverse length scale grows  logarithmically with energy used to probe 
the strings.   
This explains the linearly rising Regge trajectory 
in the Regge-pole behavior. Note that in the special case of the strings in 
1+1 dimensions, there is a longitudinal size but no transverse size, 
 and, hence,  the 
high-energy limits of the perturbative amplitudes show  power-law  behaviors, 
apart from the phase factors associated with the 
external legs.  For a discussion on the physics of the 1+1 dimensinal 
string scattering, we refer the reader to ref. \cite{jeliyo}.
}) spatial lengths 
is a natural space-time expression of the original 
Regge and resonance duality which has of course been 
the original principle, underling the 
string model (or its old name, ``dual resonance model"). 
The Regge and the resonance behaviors correspond to the 
regimes, $\Delta X_l \rightarrow \infty$ and 
$\Delta T \rightarrow \infty$, respectively. 

Let us first consider a high-energy scattering
by which we can probe the small time scale 
$\Delta T \rightarrow 0$. 
If we suppose that the uncertainty relation is nearly saturated 
in the high energy Regge-pole regime with a large  
s-channel energy and a small t-channel momentum and that the 
interaction occurs  locally along the longitudinal 
direction of string  as in the light-cone picture 
for the string, the amplitude 
is expected to be roughly proportional to $\Delta X_l \sim \ell_s^2 /\Delta T 
\propto E$ which 
implies \cite{susskind1} that the intercept $\alpha(0)$ of the leading Regge trajectory 
is 2, from the relation $E \sim E^{\alpha(t)-1} $, 
as it should be in a theory containing the graviton. 
 
In contrast to this, 
in the high energy fixed-angle scatterings with 
large s-and t-channel momenta \cite{gross}, 
we are trying to probe the region where both the time and the 
spatial scales are small. Clearly, such a region is incompatible 
with the space-time uncertainty relation. The exponential fall-off 
of the perturbative string amplitudes in this limit 
may be interpreted as a manifestation of this property. 
According to the space-time uncertainty relation, at least 
one of the two length scales must be larger than the 
string scale $\ell_s$.  

Another method of probing the small time region 
$\Delta T\rightarrow 0$ is to use black holes. 
Any finite time interval in the asymptotic space-time 
region corresponds to infinitely short time interval 
on the black hole horizons because of the infinite 
red shift. This means that in any scattering 
event of a string with a black hole, the spatial 
extension of the string at the horizon is 
effectively infinite. Thus we expect that 
the description of string-black hole scattering 
should be very different from that of ordinary 
local field theory. For example, this might imply that as for the 
observer at infinity no information loss occurs 
in the scattering with a black hole, as has been 
suggested by Susskind in the name of 
"black hole complementarity" 
\cite{susskind2}. 
Unfortunately, however, since only reliable 
discussion of string- black hole scatterings is 
through the low-energy effective field theory approximation,  
we have no satisfactory example to see how 
that difference is and what its significance is. 
In connection with this, it might be worthwhile to recall a very 
peculiar behavior of a matrix-model description for the 
$1+1$-dimensional $SL(2, R)/U(1)$ black hole, as 
has been proposed in ref. \cite{jevyoneya}.  Note that the 
behavior $\Delta X_l \rightarrow \infty$, being the 
extension in the longitudinal length, is 
valid even in $1+1$ dimensions. Our intuition 
from local field theory would lose its validity when 
we consider the string-black hole scattering, especially 
when it comes to the localization property 
\cite{pol1} of the interactions 
in the sense of target space-times. For a review on  
some controversial issues related to this problem, 
I would like to refer the reader to ref. \cite{yoneya3}.  

What is the implication of the space-time uncertainty relation in the 
large time region $\Delta T\rightarrow \infty$?
The relation tells us that we may have a small spatial scale 
$\Delta X_l \rightarrow 0$.  This is at first sight somewhat contrary to our 
intuition, since any stationary string states always have an 
extension of order of $\ell_s$. However, we can probe the 
structure of strings only through the scattering experiments, 
and therefore only observables are the S-matrix elements. 
The possibility of the non-linear $\sigma$ model approach 
shows that the asymptotic states of any string states 
can be regarded as the local space-time fields.  
It seems reasonable to interpret that the appearance of the local fields 
corresponds to the property 
$\Delta X_l \rightarrow 0$. 
This, however, does not show the possibility of directly 
probing the short spatial scales in string theory. 
The space-time uncertainty relation shows that 
doing this necessarily requires us to spent long time 
intervals, $\Delta T \rightarrow \infty$, or low energies.  
As we shall see later, the D-particles indeed provides us 
such a possibility. 

\subsection{Conformal invariance and the 
space-time uncertainty relation}

Now before proceeding to discuss the implication of the 
space-time uncertainty relation for the nonperturbative 
aspects of the string theory, we briefly describe its 
connection \cite{yoneya2} to  the world-sheet conformal invariance. 
One well known property of string perturbation theory,  
characterizing the short distance behavior of 
the string theory which is in a marked contrast to the 
local field theory, is the modular invariance. This 
shows that there is a natural ultraviolet cut-off 
of order $\ell_s^{-1}$ in string theory. The origin of the modular invariance 
can be traced back to a simple symmetry of the 
string amplitude for a world-sheet parallelogram. 
Consider the Polyakov amplitude for the mapping:  
The boundary conditions are of 
Dirichlet type :
\[
X^{\mu}(\tau, 0)=X^{\mu}(\tau, b)=\delta^{\mu 0}{A\tau\over a},
\]
\[
X^{\mu}(0, \sigma)=X^{\mu}(a, \sigma)=\delta^{\mu 1}{B\sigma\over b}.
\]
where $A, B$ are the lengths in space-time corresponding to the 
lengths $\tau, \sigma$ on the Riemann sheet using the 
conformal metric $g_{ab}=\delta_{ab}$. Then apart from a  
power behaved pre-factor, the amplitude takes the form 
\EQ
\exp [-{1\over \ell_s^2} ({A^2\over \Gamma} 
+{B^2 \over \Gamma^*})]
\EN
where 
\EQ
\Gamma \equiv {a\over b}, \, \, \Gamma^* \equiv {b\over a} ,  
\quad (\Gamma\cdot\Gamma^* =1).
\EN
Due to the conformal invariance, the amplitude 
depends on the Riemann sheet parameters only through   
the ratio $\Gamma$ or $\Gamma^*$, 
which are called the extremal length and the conjugate 
extremal length, respectively.  Clearly, the relation 
$\Gamma \Gamma^*=1$ leads to the uncertainty relation \cite{yoneya2}
\[
\Delta T \Delta X \sim 
\langle  A \rangle \langle  B\rangle \sim \ell_s^2 , 
\]
and the symmetry $(A, \Gamma) \leftrightarrow (B, \Gamma^*)$ 
is the origin of the modular invariance of a torus amplitude. 
It is only at the critical space-time dimensions 
where the conformal anomaly cancels that the 
pre-factor enjoys the required symmetry.  

This form of the amplitude is reminiscent of the 
matrix element  for a  Gaussian wave packet state $<x|g(\Delta x)>
\sim \exp -{1\over 2}({x\over \Delta x})^2$ 
saturating the Heisenberg' s uncertainty relation $\Delta x \Delta p=\hbar$ in the Wigner representation. 
\EQ
{\cal O}(x,p)\equiv 
\int \, dy \, \e^{ipy \over \hbar}\, 
<x-{1\over 2}y|\, {\cal O}_{\Delta X}\, |x+{1\over 2}y>\, 
\propto \exp -\Big[({x\over \Delta x})^2 + ({p\over \Delta p})^2\Bigr],
\EN
where ${\cal O}_{\Delta x}=|g(\Delta x)><g(\Delta x)|$ is the 
density operator representing the 
Gaussian state. The correspondence,  
$(\lambda, A, B, \sqrt{a/b}, \sqrt{b/a}) 
\leftrightarrow (\hbar, x, y, \Delta x/\sqrt{\hbar}, \Delta p/\sqrt{\hbar})$, 
suggests that we integrate over the uncertainty $\Delta x$ 
in analogy with the integration over the modular parameter 
in string theory.  Thus the statistical density operator 
$\rho \equiv \int d(\Delta x) \, {\cal O}_{\Delta x}$ 
is an analogue of the string amplitude.

\section{D-particle dynamics and the space-time uncertainty 
relation}

Let us now discuss the possible relevance of our 
space-time uncertainty relation to the nonperturbative 
aspect of the string theory. 
One of the most important observation in the 
recent development is that 
the Dirichlet branes are solitonic excitations  
with Ramond-Ramond charge \cite{pol2} which are crucial ingredients 
for the validity of the S-duality of string theory. 
The dynamics of the D-branes are 
described by the collective degrees of freedom
 in terms of open strings with Dirichlet 
boundary conditions. The collective coordinates are 
matrices living on the D-branes and associated with 
the open-string vertex operators. The matrix nature 
of the collective degrees of freedom comes from the Chan-Paton 
factor of the open-string vertex operators. 

For example, for the 
D0-branes (or D-``particles") in the type IIA 
theory, the collective variable 
is  one-dimensional Hermitian matrices $X^a_{ij}(t)$ 
and their fermionic partners,  
where $t$ is the time along the world line of the 
D-particle and the space index $a$ runs from 1 to 9. 
. The Chan-Paton indices $i, j$ run from $1$ to 
$N$  with N being the number of the D-branes involved. 
The effective 
action \cite{witten1} for the collective fields $X^{\mu}_{ij}(t)$ is 
nothing but the dimensional reduction 
of $10$ dimensional $U(N)$ super Yang-Mills theory to 
$0+1$ dimensions at least for sufficiently low energy 
phenomena: 
\EQ
S=\int dt 
{1\over 2g_s}\Tr\biggl\{ \dot{ X^a}\dot{ X^a} 
 - {1\over 2}[X^a,X^b]^2 
 + fermions\biggr\}. 
\label{D0effectiveaction}
\EN
Here we used the unit $\ell_s =1$ for simplicity 
and assume the $A_0=0$ 
gauge. This effective action clearly shows that the system 
along the classical flat direction $[X^a,X^b]=0$ corresponds to 
N free nonrelativistic particles of mass $1/g_s$ 
whose coordinates are the eigenvalue of the 
matrices $X^a$.  

Is it possible to probe arbitrary 
short distances by a scattering experiment 
using these D-particles? \cite{kp}$^,$\cite{douglasetal} The interaction of the 
D-particles are governed by the 
off-diagonal elements of the matrices which corresponds to the lowest 
modes of the open strings stretching between the D-particles. 
The lowest frequency of the off-diagonal elements $X^{(\mu)}_{ij}$ is 
of order $|\lambda_i -\lambda_j|$ where the 
$\lambda_i$ is the coordinate of the $i$-th D-particle. 
This means that the characteristic time scale 
$\Delta T \sim {1\over |\lambda_i -\lambda_j|}$ increases as 
we try to probe shorter distances. This is just the property 
required by the space-time uncertainty relation. 
We can then expect that a scattering of very slow D-particles 
would make possible to probe very short distances. 
However, quantum mechanics tells us that 
if we spent too long time, the wave packets of the 
D-particles spread and wash out the information 
on the short distances. The spreading of the wave packet 
is smaller if the mass of the particles are heavier. 
In our case,  the mass of the D-particles is proportional to the 
inverse string coupling $1/g_s$.   Thus there must be a 
characteristic spatial scale which gives the order of the 
shortest possible distances probed by the D-particle 
scattering. This is easily obtained by the scaling argument. 
By redefining the time and the coordinate matrices as 
\EQ
X =g_s^{1/3}\tilde{X}, \,  \, \, t =g_s^{-1/3}\tilde{t}, 
\EN
we can completely eliminate the string coupling from the 
effective action. Thus we see that the characteristic 
spatial and time scales, restoring the 
string scale, are given by 
$g_s^{1/3}\ell_s$ and $g_s^{-1/3}\ell_s$, respectively. 
As is well known, the spatial scale $g_s^{1/3}\ell_s$ is 
nothing but the 
11 dimensional Planck scale of the M-theory 
\cite{witten2}. On the other hand, 
the time scale is given by the inverse power $g_s^{-1/3}\ell_s$ 
as is required by the space-time uncertainty relation. 
We note that for the validity of the above 
dimensional arguments, the existence of the supersymmetry 
is important, since if the supersymmetry did not 
exist the zero-point oscillation would generate a long-range 
potential for the eigenvalue coordinates and 
would ruin the simple scaling property of the 
classical effective action. 

These results are readily derived, without explicitly using the 
effective action, if one first assumes  
the space-time uncertainty relation as an underlying 
principle of the string theory including nonperturbative 
objects \cite{liyo}. 
Consider the scattering of two D-particles of mass $\ell_s/g_s$ with 
the impact parameter of order $\Delta X$ and the relative 
velocity $v$ which is assumed to be much smaller than the 
light velocity. Then the characteristic interaction time $\Delta T$ is 
of order ${\Delta X\over v}$. Since the impact parameter is of the 
same order as the longitudinal length of the open strings 
mediating the interaction of the D-particles, we can 
use the space-time uncertainty relation in the form
\[
\Delta T \Delta X \sim \ell_s^2 \Rightarrow 
{(\Delta X)^2 \over v} \sim \ell_s^2 
\]
This gives the order of the magnitude for the possible 
distances probed by the D-particle scatterings with 
velocity $v \, (\ll 1)$. 
\EQ
\Delta X\sim 
\sqrt{v}\ell_s.
\EN
The same result is obtained from the effective low-energy action 
by using the Born-Oppenheimer approximation \cite{douglasetal} 
for the coupling between the diagonal and off-diagonal 
matrix elements. 
The spreading of the wave packet of the D-particle 
during the time interval $\Delta T \sim {\ell_s\over \sqrt{v}}$ 
is easily estimated as 
\[
\Delta X_w \sim 
\Delta T\Delta_w v \sim 
{g_s\over v}\ell_s 
\]
where $\Delta_w v \sim g_s {1\over \sqrt{v}}$ is the 
uncertainty of velocity caused by the ordinary uncertainty relation 
for a nonrelativistic particle with energy uncertainty,  
$\Delta E \sim 1/\Delta T \sim {\sqrt{v}\over \ell_s} 
\sim {v\Delta_w v \ell_s\over g_s}$. 
The condition 
$
\Delta X_w  (=g_s {\ell_s\over v}) < \Delta X  
(=\sqrt{v}\ell_s )
$
leads 
\EQ
\Delta X >  g_s^{1/3}\ell_s . 
\EN

Actually, by considering the systems with many 
D-branes we 
can probe the smaller spatial scales than the 11 dimensional 
Planck length. For example, when we consider a D-particle in the 
presence of many (=$N$) coincident D4-branes, the effective 
kinetic term for the D-particle is given by \cite{douglasetal}
\EQ
S_{{\rm eff}}= \int dt [{1\over 2g_s\ell_s}(1+{Ng_s\ell_s^3 \over  
r^3})v^2 +  {\cal O}(Nv^4 \ell_s^6/r^7)]
\label{D0D4action}
\EN
where $r$ is the distance between the D-particle and the D4-branes. 
When the distance $r$ is much shorter than $(Ng_s)^{1/3}\ell_s$,
the effective mass of the D-particle is given by
$m\sim N\ell_s^2/r^3 \gg {1\over g_s\ell_s}$. We can then easily  
check that  the spread of the D-particle wave packet can be neglected
during the time $t \sim v^{-1/2}\ell_s$, compared with the scale 
determined above from the consideration of the 
space-time uncertainty relation 
$\sqrt{v}\ell_s$ for large $N$. 
This allows us to probe arbitrary  
short lengths
with respect to the distance between the D-particle and D4-branes 
and hence to see even the singular nature of the effective metric 
in the action (\ref{D0D4action}).   
Since the time scale grows indefinitely, however, 
we cannot talk about the interaction time in any meaningful way
in the limit of short spatial distance. 

These considerations indicate, I believe,  the universal nature 
of the space-time uncertainty relation as one of the possible 
principles underlying the nonperturbative string theory. 

As is well known now, an exciting interpretation 
for the effective action (\ref{D0effectiveaction}) is 
to regard it as the {\it exact} action of the M-theory for the 9 (=11-2) 
transverse degrees of freedom in the infinite-momentum 
frame in 11 dimensions (so called `M(atrix) theory')
\cite{banksetal}.  
This ansatz has already passed a number of  nontrivial 
tests. 
In the 11 dimensional M-theory which is 
believed to be the underlying theory of the type IIA 
string theory, the D0-brane is interpreted as the 
Kaluza-Klein excitation of the 11 dimensional 
massless fields.  The compactification radius of the 
11th dimension is $g_s\ell_s$, and hence the unit of the Kaluza-Klein 
momentum in 11 dimensions is just the mass of D0-branes. 
The fundamental assumption is 
that the momentum of the D0-brane in the 
11th direction is always equal to this 
single unit of the momentum and hence the total number $N$ 
of the D0-branes is just proportional to the 
total 11th momentum. Thus going to the infinite momentum 
frame is equivalent to the large $N$ limit. 
Combined our discussion with this proposal, it seems 
natural to try to construct possible covariant formulations 
of the M-theory by taking account the space-time uncertainty 
relation.  My own attempt toward this direction has  so far 
been unsuccessful. 
The covariant formulation necessarily requires all of the 
space-time coordinates as matrices. 
The crucial step would thus be to identify a natural higher symmetry 
such that it allows us to go to the light-cone gauge effective 
action (\ref{D0effectiveaction}) in which the light-cone time is  
reduced to a single variable  and the total $11$th momentum, 
 measured by the basic unit $1/g_s$, 
becomes  equal to $N$.  Hence, necessarily, 
the order of the matrices itself becomes 
a dynamical variable. 
It turns out that this is very nontrivial to implement 
in the framework of the matrix models.  
From physical side, one of the difficulty is that 
in the ordinary Lorentz frame, we 
have perhaps to treat D-particles and anti-D-particles simultaneously. 
However, at present, almost nothing is known concerning 
about the brane-anti-brane interactions, except that   
the usual perturbative description in terms of the open strings 
breaks down near the string scale \cite{bankssuss}. 
In order to achieve the covariant formulation,  
we would have to make some acrobatic twists to the ordinary 
matrix model.  

\vspace{0.5cm}
\noindent
{\it Remarks} 
\begin{enumerate}
\item The above scaling argument can be extended to Dp-branes. 
The longitudinal (including the time direction) and transverse lengths always scale 
oppositely: The powers are  $g_s^{-1/(3-p)}$ and 
$g_s^{1/(3-p)}$, respectively. This reflects the fact that the interaction 
of D-branes are mediated by open strings satisfying the 
space-time uncertainty relation for the time and the longitudinal lengths of the 
open string. 
Note that the longitudinal direction of the 
open strings corresponds to the transverse 
directions for the D-branes. 
For example, for $p=1$ (D-string), the 
characteristic spatial scale is $g_s^{1/2}\ell_s$. 
The meaning of this scale is not yet fully understood. 
In any case (provided $p\ge 0$), 
we see that the space-time uncertainty relation 
is satisfied, although the characteristic spatial and time scales  
vary for each specific case. 
The case $p=-1$ (D-instanton) is very special 
in the sense that all the space-time directions 
are `transverse' and there is no time 
evolution;  but we shall argue, 
in the next section,  
that this case is also consistent with the 
space-time uncertainty relation. 
\footnote{
As for a possible interpretation of the string-D-instanton interactions 
\cite{klebthor} from the viewpoint of the space-time 
uncertainty relation, see \cite{liyo}. Essentially, the instanton-string  interaction is necessarily in the regime $\Delta T\rightarrow 0$, and correspodingly is governed by the massless long-range 
exchange interactions $\Delta X \rightarrow \infty$.} 
\item On the other hand, if we go to the strong coupling regime 
of the string theory, we expect that the D-branes become the 
lighter excitation modes than the fundamental strings. 
 In particular, in the type IIB 
theory, the conjectured S-duality relation says that the 
role of the fundamental strings and the D-strings are 
interchanged in a completely symmetrical way. 
Then the main interaction among the D-strings 
should be through the ordinary splitting and merging  
of the D-strings. Since the string scale is invariant 
under the S-duality transformation $g_s \leftrightarrow 1/g_s$, 
we expect that our space-time uncertainty relation is 
satisfied even in the strong coupling limit of the type IIB theory. 
Thus it is plausible that the uncertainly relation is universally 
valid independently of the strength of the string coupling, although 
it is tested only in the weak coupling string theory. 
It should however be expected that in the 
intermediate coupling regions, 
it would be difficult to saturate the uncertainty relations 
since in this region the halo around the D-branes (F-strings) consisting 
of the cloud of the F-strings (D-brane) should be generally 
important around the string scale.  
\end{enumerate}

\section{Type-IIB matrix models as a possible 
realization of the space-time uncertainty relation}

\subsection{D-instanton and the type IIB matrix model}
The D0-brane, D-particle, is the lowest dimensional 
excitation of the type IIA superstring theory 
which allows only even dimensional D-branes. 
The type IIB theory, on the other hand, allows only 
odd dimensional D-branes. In particular, the object 
of the lowest dimension is the D-1-brane, namely, 
the D-instanton with $p=-1$. For the D-instanton, all of the 
space-time directions, including even the time direction, 
becomes the transverse directions. 
The interaction of the D-instantons are therefore 
described by the open strings which obeys the 
Dirichlet boundary condition with respect to all the 
directions. Hence the effective action for the 
D-instantons, at least for 
low-energy phenomena, are described by the $10$ dimensional 
super Yang-Mills theory reduced to a space-time point: 
\EQ
S_{{\rm eff}}= -{1\over g_s}\Tr ([X_{\mu}, X_{\nu}])^2 + 
{\rm fermionic 
\, \, \, part}
\EN
which is covariant from the beginning.  We again suppress 
the string scale by taking the unit $\ell_s=1$. 
The D-instanton is rather special in the sense that 
the only ``physical" degrees of freedom  are the lowest massless modes 
of the open strings. The reason is that only 
on-shell states satisfying the Virasoro condition 
are the zero-momentum ``discrete" states 
of the Yang-Mills gauge fields which are 
identified with $X_{\mu}$'s and their 
super partners.  
Other infinite number of the 
 massive degrees of freedom of the open strings 
are excluded as the solution of the Virasoro condition 
and therefore should be regarded as a sort of 
auxiliary fields. 
We then expect that the 
exact effective action for them should be 
expressible solely in terms of the collective fields which 
are nothing but the lowest modes $X_{\mu}$. 
If we further assume that all the higher dimensional 
objects including the fundamental strings could be various 
kinds of  bound states or collective 
modes composed of infinitely many
D-instantons,
\footnote{
For example, in usual gauge field theories, 
a monopole can formally be interpreted  as being 
composed of infinite number of  instantons along a line. 
}  we may proceed to postulate that 
a matrix model defined at a single point 
might give a possible nonperturbative 
definition of the type IIB string theory.  
We however note that this is a very bold assumption: 
For example, unlike the case of the M(atrix) theory, 
here there is no rationale why we can neglect 
the configuration in which both instantons and 
anti-instantons are present. If we can only hope that 
suitable choice of the model might 
take into account such effect implicitly.

A concrete proposal along this line  was 
indeed put forward first by Ishibashi, Kawai, Kitazawa and Tsuchiya 
\cite{ikkt}. 
Their proposal is essentially to regard the above 
simple action as the exact action in the following form. 
\EQ
Z_{\rm{IKKT}}= \sum_N \Bigl(\prod_{\mu}\int \, d^{N^2}X_{\mu}
\Bigr) d^{16}\psi
\exp S[X, \psi, \alpha, \beta]
\EN
\EQ
S[X, \psi, \alpha, \beta] \equiv  -\alpha N +
\beta \Tr_N{1\over 2}[X_{\mu}, Y_{\nu}]^2
- \cdots
\label{ikkt}
\EN
with $\alpha \propto {1\over g_s\ell_s^4}, \beta \propto {1\over g_s}$ 
and the summation over $N$ is assumed.  
 Indeed, it has been shown that the model 
reproduces the long-distance behavior of 
D-brane interactions expected from the 
low-energy effective theory, the type IIB supergravity in 10 dimensions. 
Very recently, they have presented further 
arguments \cite{fkkt} which suggest that the model 
can reproduce the light-cone string field theory.

\subsection{On the Schild action formulation of string theory}
Let us now discuss the connection of the 
space-time uncertainty relation with the  type IIB models of this type. 
Ishibashi et al motivated their proposal by  the formulation of 
Green-Schwarz superstring action in the Schild gauge 
\cite{schild}. 
In fact, the formulation of the string theory in the 
Schild gauge \cite{schild} is very natural from our viewpoint. 
So, let me begin from a discussion of the Schild action. 
\EQ
S_2 =-\int d^2\xi\, e \Bigl\{{1\over e^2}
[-{1 \over 2\lambda^2}(\epsilon^{ab}\partial_a X^{\mu}\partial_b  X^{\nu})^2] + 1
\Bigr\} +\cdots
\label{Schildaction}
\EN
where $e=e(\xi)$ is an auxiliary field with the same transformation 
property with the volume density of the world sheet 
and $\lambda=4\pi\alpha'$. 
The classical equation of motion for the string coordinate takes 
the same form as that from the Nambu-Goto action 
which is obtained from $S_2$ by eliminating the 
auxiliary field $e$. 
Alternatively, we can also derive the Polyakov action directly 
as follows. 

The  action can be made quadratic by introducing three  
auxiliary fields $t^{12}=t^{21}, t^{11}, t^{22}$, which forms 
a tensor density of weight two, as 
\EQA
S_2 &=& -\int d^2\xi \, {1\over \lambda^2 e}
[{\dot X}^2{\acute X}^2 - ({\dot X}\cdot {\acute X})^2] -\int d^2\xi 
\, e
\nonumber \\
&\equiv & S_(t, e, X)  - \int d^2\xi  {1\over e}\det \tilde{t}
\label{gaussiancompletion}
\EQN
where 
\EQ
S(t,e,X)= \int d^2\xi \, {1\over e}[\det t 
+{1\over \lambda} t^{ab}\partial_aX \cdot \partial_bX ] -\int d^2\xi \, e,
\EN
and we have defined the shifted auxiliary fields $\tilde t^{ab}$ by 
\EQ
\tilde{t}^{11}= t^{11}+ {1\over \lambda}{\acute X}^2, 
\quad \tilde{t}^{22} \equiv \tilde{t}^{21}=t^{22}+ {1\over \lambda}{\dot X}^2, \quad \tilde{t}^{12}=
t^{12}- {1\over \lambda}{\dot X}\cdot {\acute X}. 
\EN
The new action $S(t,e,X)$ containing the auxiliary field $t^{ab}$ 
is quantum-mechanically equivalent to the original Schild action 
since the difference is a free quadratic form 
$\int d^2\xi {1\over e}\det \tilde{t}$ with respect to 
the auxiliary fields $\tilde{t}^{ab}$, 
apart from an induced ultra-local measure   
which we included in the definition of the measure for $e$ :
\EQ
Z_S\equiv \int {[dX][de]\over [d(diff_2)]} \, \e^{-S_2} 
\propto \int {[dX][de][dt]\over [d(diff_2)]} \, \e^{-S(t,e,X)}.
\EN
Let us then make a change of variables 
$$t^{ab}\rightarrow g^{ab}, 
e \rightarrow \tilde{e}$$
where $g_{ab}$ transforms 
as the standard world-sheet metric and $\tilde{e}$ as
a scalar,  
$ t^{ab}= g^{ab}e^2, e = \tilde{e}\sqrt{g}$,  
and get 
\EQ
Z_S  \propto \int {[d\tilde{e}] [dg][dX] \over [d(diff_2)]}
\exp \Bigl(
-\int d^2 \xi (\tilde{e}^3-\tilde{e}) \sqrt{g}
-
{1\over \lambda}\int d^2\xi \tilde{e}\sqrt{g}g^{ab}
\partial_aX\cdot \partial_bX
\Bigl). 
\label{zexg}
\EN

At this point, we can assume that the measure 
in the partition function $Z_S$ is defined 
 such that the total integration measure 
$[d\tilde{e}][dX][dg]$ obtained in these transformations is  
reparametrization invariant. 
We then decompose the measure $[dg]$ as usual 
$$
{[dg]\over [d(diff_2)]}=[d{\rm Weyl}]{[dg] \over [d{\rm Weyl}] [d(diff_2)]}.
$$
The functional metrics for the Weyl modes and auxiliary scalar fields $\tilde{e}$ 
are, in the conformal gauge $g_{ab}=\delta_{ab}\, \e^{\phi} $ for simplicity, 
\EQ
\| \delta (\e^{\phi})\|^2 =\int d^2\xi \e^{-\phi}(\delta \e^{\phi})^2,
\label{phimetric}
\EN
\EQ
\|\delta\tilde{e}\|^2 =\int d^2\xi \e^{\phi}(\delta\tilde{e})^2,
\label{tildeemetric}
\EN
respectively.  Note that in the former expression we have 
chosen the integration variable to be the density itself $\e^{\phi}=\sqrt{g}$. 
Now, when the conformal anomaly cancels, there is no induced 
kinetic term arising from the integration measure 
$[dg][dX]/[d{\rm Weyl}][d(diff_2)]$  for the 
Weyl mode in the world-sheet metric $g_{ab}$ 
which therefore completely decouples from the second term 
of the action in the expression (\ref{zexg}). 
Then, the integration over the Weyl mode gives a $\delta$-function  
constraint $\tilde{e}=1$. 

Here we assumed  that the integration 
measure  in the partition function is chosen such 
that  the diffeomorphism invariant 
functional $\delta$-function formula  is valid, 
$$
\int [d\tilde{e}][d\sqrt{g}]f[\tilde{e}]\exp (\int d^2\xi \sqrt{g}\tilde{e})=f[0], 
$$
for arbitrary $f[\tilde{e}]$ which is independent of $\sqrt{g}$ as a 
functional of a scalar field $\tilde{e}$. Formally, this amounts to 
postulating that the integration measure is 
defined such that the factors $\e^{-\phi}$, $\e^{\phi}$ 
in the functional metrics 
(\ref{phimetric}) and (\ref{tildeemetric}), respectively, cancel 
each other in going from the original functional metrics (\ref{phimetric}) and 
(\ref{tildeemetric}) to 
$\overline{\| \delta (\e^{\phi})\|}^2 \equiv \int d^2\xi (\delta (\e^{\phi}))^2$ and 
$\overline{\|\delta\tilde{e}\|}^2\equiv \int d^2\xi (\delta\tilde{e})^2$, 
and that the integration range for $\sqrt{g}\equiv \e^{\phi}$ 
is in fact extended to the whole imaginary axis. 
The first assumption
 can be justified by adopting the point-splitting regularization scheme 
for the formal Jacobian $J=|\det (\e^{\phi(\xi_1)-\phi(\xi_1')
+\phi(\xi_2)-\phi(\xi_2')}\delta(\xi_1, \xi_1';\xi_2, \xi_2')|
$ associated with the 
transition 
$[d\tilde{e}][d\sqrt{g}] =\overline{[d\tilde{e}][d\sqrt{g}]}\, J$, 
such that the sets of the discretized world-sheet points arising 
from the two factors $\e^{-\phi}$, $\e^{\phi}$ are the same, 
$\{\xi_i\}=\{\xi_i'\}$.  The $\delta$-function 
$\delta(\xi_1, \xi_1';\xi_2, \xi_2')$ can be regularized by, e.g., 
the reparametrization invariant heat kernel method.  
The assumption for the integration range  is not inconsistent  
since the fields $g_{ab}$ come from the 
auxiliary fields $t^{ab}$.  Also we have discarded the 
nonsensical solution $\tilde{e}=0$ which can be excluded 
by making a justifiable assumption that the measure for $\tilde{e}$ vanishes at 
$\tilde{e}=0$. Note that the Gaussian integration over the 
shifted field $\tilde{t}^{ab}$ in (\ref{gaussiancompletion}) 
indeed produces such a factor. 

Thus the final result is, apart from a proportional numerical 
constant, that 
\EQ
Z_S =\int {[dX][dg]\over [d(diff_2)] [d{\rm Weyl}]} \,\e^{-S^e_P}
\EN
in critical space-time dimensions where $S_P^e$ is the 
 Euclidean Polyakov action. 
Obviously, the argument goes through for 
superstrings by including necessary fermionic variables 
once the additional terms for the 
bosonic Schild action to ensure the supersymmetry are  
determined. 
Although the above argument is not rigorous, 
it seems now fairly clear that we have the quantum string 
theory using the Schild action, which is equivalent to the standard 
formulation based on the Polyakov action.

\subsection{The space-time uncertainty relation form the 
Schild action}
The Schild action has no manifest Weyl (or conformal) invariance. 
Nonetheless, it is equivalent with the Polyakov action 
when the conformal anomaly cancels. 
Where is then the conformal structure hidden?
It is easy to see that the Virasoro constraint for the canonical 
variables
\EQA
{\cal P}^2 + {1\over 4\pi\alpha'}{\acute X}^2 &=& 0 ,\label{virasoro1}\\
{\cal P}\cdot {\acute X} &=&0 .\label{virasoro2}
\EQN
is obtained  after using the 
constraint coming from the variation with respect to the 
auxiliary field:
\EQ
 {1\over e}\sqrt{-{1\over 2}(\epsilon^{ab}\partial_a X^{\mu}\partial_b  X^{\nu})^2}=\lambda .
\label{conformalconstraint}
\EN
 Since the Virasoro condition expresses the 
conformal invariance of the world-sheet field 
theory, we can say that the conformal invariance of the string theory 
is encoded in the condition (\ref{conformalconstraint}) in the 
Schild-gauge formulation. 
For this reason, we call the condition (\ref{conformalconstraint}) 
the ``conformal constraint". 
On the other hand, we have already emphasized that our 
space-time uncertainty relation can be regarded as 
a direct consequence of the conformal invariance. 
Thus it is natural to expect that the constraint 
(\ref{conformalconstraint}) is closely related to the 
spacetime uncertainty relation. 
Indeed, using the known relation 
\cite{hoppe} between the 
commutator for general Hermitian $N\times N$ matrices 
and the Poisson bracket in two-dimensional phase space 
$(\tau, \sigma)$ in the large $N$ limit, we 
can make the following correspondence,
\EQ
\{X_{\mu}, X_{\nu}\} \equiv {1\over e}
\epsilon^{ab}\partial_a X_{\mu}\partial_b X_{\mu}
\leftrightarrow [X_{\mu}, X_{\nu}] .
\EN
\EQ
\Tr \, (\cdots) \leftrightarrow  \int d\tau d\sigma e \, \, (\cdots) .
\EN
Then the constraint coming from the Schild action is 
interpreted as the one for the 
commutator of the matrices 
\EQ
-{1\over 2}([X^{\mu}, X^{\nu}])^2=\lambda^2 I
\label{quantumconstraint}
\EN
This leads to the inequality in the Minkowski metric
\EQ
\langle ([X^0, X^i])^2\rangle  \, \, \ge \lambda^2, 
\EN
which is just consistent with the uncertainty relation
for the time and spatial lengths. This seems to be the simplest possible 
expression of the space-time uncertainty relation 
in a way compatible with Lorentz covariance \cite{yoneya4}. 

\subsection{Microcanonical matrix model as a  
realization of the space-time uncertainty relation}

It is now reasonable to try to construct some 
matrix model which takes into account the condition 
(\ref{quantumconstraint}). In the following, we discuss   
a provisional model \cite{yoneya4} which leads to the 
model (\ref{ikkt}) at least for low-energies. 

We require that the condition (\ref{quantumconstraint}) 
 in a weaker form, namely, as an average 
\EQ
\langle {1\over 2}([X_{\mu}, X_{\nu}])^2 \rangle \equiv 
\lim_{N\rightarrow \infty}
{1\over N}\Tr {1\over 2}([X_{\mu}, X_{\nu}])^2=-\lambda^2 .
\label{matrixqcondition}
\EN
Here, $\langle \cdot\rangle$ denotes the expectation value 
with respect to the $U(N)$ trace as indicated, and 
the large $N$ limit is assumed to include the case of arbitrary 
number of D-instantons. The large $N$ limit is also necessary  
in order to include, for instance,  a static configuration of the 
D-string with the relation 
\[
[X_0, X_i] \propto I
\]
for some spatial direction $i$. 
The  fundamental partition function is then defined as 
\EQ
Z=
\int \Bigl(\prod_{\mu=1}^{10} \, d^{\infty^2}X_{\mu}
\Bigr){\cal J}[X]
\delta(\langle {1\over 2}([X_{\mu}, X_{\nu}])^2 \rangle+\lambda^2)
\EN
where $d^{\infty^2}X_{\mu}$ is 
the large N limit of the 
standard $U(N)$ invariant Haar measure and  
the factor ${\cal J}[X]$ is an 
additional measure factor to be determined below. 

Obviously, our model is similar to a microcanonical partition 
function of classical statistical mechanics. In the latter case, 
we adopt the Liouville measure of the phase space, appealing 
to the Liouville theorem and ergodic theory. In our case, 
based on the interpretation of the instanton collective 
coordinates, we require the cluster property that the 
effective dynamics for clusters of instantons which are 
separated far apart from each other should be 
independent.  The configuration of separated clusters 
are represented by diagonal blocks of smaller matrices 
$Y_{\mu}^{(a)}  \, \, (N_a\times N_a)$ 
and $Y_{\mu}^{(b)} \, \, (N_b\times N_b)$ 
embedded 
in the original large $N\times N$ matrices $X_{\mu}$ and the 
distance between two clusters are measured by the 
the difference of the center of mass coordinates 
$\ell_{a, b}\equiv |{1\over N_a}\Tr_a Y_{\mu}^{(a)} 
-{1\over N_b}\Tr_b Y_{\mu}^{(b)}|$. 
In general, the fluctuations of the off-diagonal elements 
generate effective long range interaction of the form 
$\log \ell_{a,b}$ coming from the Vandelmonde measure 
of the matrix integral. This necessarily violates the 
cluster property. A natural way out of this problem is 
to require supersymmetry for the partition function by 
introducing the fermionic partner to the collective 
coordinates to cancel the Vandelmonde measure. 
 Of course, if one tries to derive the 
effective theory for D-instantons from the 
type IIB string theory, one would have such fermionic 
partners, automatically. In this way, we are led to introduce the 
ansatz for the measure factor ${\cal J}[X]$
\EQ
{\cal J}[X] = \int d^{16}\psi' \exp ({1\over 2}\langle \overline{\psi}
\Gamma_{\mu} [X_{\mu}, \psi]\rangle)
\EN
where $\psi$ is the hermitian 
$N\times N$ matrix whose elements are Majorana-Weyl 
spinors in 10 dimensions and the prime in the integration volume 
denotes  that possible fermion zero-modes should be removed for the 
partition function. 
The supersymmetry is easily established after  rewriting  
the partition function by introducing an auxiliary constant multiplier $c$, 
\EQ
Z=\int \, dc \, \Big(\prod_{\mu}d^{\infty^2}X_{\mu}\Bigr)
d^{16}\psi'  
\exp\Bigl[
c(\langle {1\over 2}([X_{\mu}, X_{\nu}])^2 \rangle+\lambda^2) + 
{1\over 2}\langle \overline{\psi}
\Gamma_{\mu} [X_{\mu}, \psi]\rangle\Bigr] .
\label{microcanonicalmtheory}
\EN 
The action in this expression is  invariant under the 
two (global) supersymmetry transformations 
\EQA
\delta_{\epsilon}\psi&=&ic [X_{\mu}, X_{\nu}] \Gamma_{\mu\nu}\epsilon 
\\
\delta_{\epsilon}X_{\mu}&=&i\overline{\epsilon}\Gamma_{\mu}\psi 
\\
\delta_{\epsilon}c&=&0 \\
\delta_{\eta} \psi &=& \eta \\
\delta_{\eta} X_{\mu}&=&0 \\
\delta_{\eta}c&=&0
\EQN 
where the Grassmann spinorial parameters $\epsilon, \eta$ are ``global", 
i.e., the 
unit matrix with respect to the $U(N)$ indices. 
These symmetries can also be derived by the similarity with the 
Green-Schwarz formulation \cite{greenschwarz} 
of superstrings with a special gauge condition for the local 
$\kappa $ symmetry.  
The presence of the supersymmetry ensures that the 
classical configurations which preserves a part of the 
supersymmetry can enjoy the BPS property. 
For example, if  the commutator $[X_{\mu},X_{\nu}]$ 
is proportional to the unit matrix, there remains just 
a half of the supersymmetry.  

We emphasize that our argument is still provisional. 
For instance, we cannot exclude the 
possibility of lower dimensional space-times, since we 
can construct the similar models as the dimension reductions of any  
supersymmetric Yang-Mills theories. 
However, the 10 dimensions is the largest dimension 
where the above construction works and only 
in this dimension the correct long range forces 
for D-strings emerges. 

Now from our viewpoint, 
the IKKT model can be naturally derived as the effective 
low-energy theory of many distant clusters of D-branes. 
We introduce the block-diagonal matrices whose 
entries are the smaller $N_a\times N_a$ matrices $Y_{\mu}^{(a)}$ and 
require the conditions 
\EQ
-\sum_{a=1}^n \Tr_a {1\over 2}[Y_{\mu}^{(a)}, Y_{\nu}^{(a)}]^2 =N\lambda^2,
\label{condition1}
\EN
\EQ
\sum_{a=1}^n \, N_a =N \, \, (\rightarrow \infty).
\label{condition2}
\EN
Now let us suppose to evaluate the fluctuations around 
the backgrounds by setting $X_{\mu}=Y_{\mu}^b+ \tilde{X}_{\mu}$. 
In the one-loop approximation,
 the calculation is entirely the same\footnote{
Note that the one-loop calculation is  
reduced to the determinant in both models.}
 as the 
IKKT model \cite{ikkt}, and it is easy to see that the 
leading order behavior of the interaction contained in the  
one-loop effective action for the backgrounds 
decreases as $O({1\over \ell_{a,b}^8})$ in the limit of 
large separation.  Remember  that here again the supersymmetry is 
crucial. 
The distant D-brane systems therefore can be treated as independent 
objects in this approximation, and hence we can take into account the 
conditions (\ref{condition1}), (\ref{condition2}) in a 
statistical way by introducing two Lagrange 
multipliers $\alpha$ and $\beta$.  
Then the effective partition function $Z_{\rm{eff}}$ for the D-brane 
subsystems described $Y_{\mu}$ within the 
semi-classical approximation is given by the grand canonical 
partition function, 
\EQ
Z_{\rm{eff}}= \sum_N \Bigl(\prod_{\mu}\int \, d^{N^2}Y_{\mu}
\Bigr) d^{16}\psi
\exp S[Y, \psi, \alpha, \beta],
\label{grandcanonicalform}
\EN
\EQ
S[Y, \psi, \alpha, \beta] \equiv  -\alpha N +\beta \Tr_N{1\over 2}[Y_{\mu}, Y_{\nu}]^2
-{1\over 2}\Tr \, \overline{\psi}[\Gamma_{\mu}, Y_{\mu}]\psi ,
\EN
where by $N$ we denote the order of the background submatrix $Y$ 
and $\Tr_N$ is the corresponding trace.  This 
form is identical with the IKKT model and 
explains the origin of the two 
parameters.  
In our case, however, we have the condition 
\EQA
-\langle\langle  {1\over 2}[Y_{\mu}, Y_{\nu}]^2 \rangle\rangle 
&\equiv& -{\sum_N \Bigl(\prod_{\mu}\int \, d^{N^2}Y_{\mu}\Bigr) d^{16}\psi
\, {1\over N}
\Tr_N {1\over 2}[Y_{\mu}, Y_{\nu}]^2 \exp S[Y, \psi, \alpha, \beta]
\over \sum_N \Bigl(\prod_{\mu}\int \, d^{N^2}Y_{\mu}\Bigr) d^{16}\psi 
\exp S[Y, \psi, \alpha, \beta]} \nonumber \\
& = & \lambda^2
\label{condition3}
\EQN
ensuring the quantum constraint by which 
 the Lagrange multipliers should in principle be determined.   
 The original microcanonical model has two parameters, string 
scale $\sqrt{\lambda}$ and $N$ 
which corresponds to two parameters $\alpha, \beta$ 
of the low-energy effective action. In the large $N$  limit,  
we expect there remains a dimensionless  parameter other than the 
fundamental length. It should be related to 
 the  coupling constant of the effective string theory. 
In the nonperturbative string theory, we expect that 
the coupling constant should be 
 determined dynamically from the vacuum expectation values of the 
scalar background fields such as dilaton and/or  
its dual partner, scalar axion.   
It would be very interesting to see 
whether the condition (\ref{condition3}) is 
consistent with this expectation. For example, 
the form (\ref{condition3}) can be interpreted as the 
effective dilaton equation of motion 
\[
{\partial \over \partial \phi} Z=0
\]
if $\alpha \sim {1 \over g_s\ell_s^4}, \beta \sim {1\over g_s}$. 
As is discussed in ref. \cite{ikkt}, this is just the correct 
parameter dependence required 
to reproduce the classical D-string interactions. 

\section{Discussions}

There are many reasons that the discussions 
given here for the relevance and formulation of the 
space-time uncertainty principle for the string theory 
are yet provisional and should be 
regarded only as a motivation for further exploration 
of the basic principles behind and beyond the string theory. 

Among others, one of the most crucial puzzle  seems to me  
 why the gravity is contained 
in the matrix model formulation of string theory. 
Both in the type IIA and IIB models, we start from the 
flat space-time background and there is, 
at least apparently, no symmetry which can be 
connected to general coordinate transformation and 
its supersymmetric generalization. Of course, there exists 
the space-time supersymmetry which is maximal in each case. However, 
it is not at all clear how it must be elevated to local 
symmetry. It is possible that such 
local symmetries are hidden in some form in the 
the $SU(N)$ symmetries in the large $N$ limit. 
In that case, our usual understanding of the 
background space-time should be 
reconsidered for the matrix-model approaches. 
We need some correspondence principle which 
relates the classical geometric field theory to 
matrix models containing gravity. 

Another very important question is how to unify the entirely different 
realizations of the space-time uncertainty relation in 
the D-particle (type IIA) and D-instanton (type IIB) cases. 
In the D-particle case, the time variable is treated just as the  
ordinary time in quantum mechanics. The space-time 
uncertainty relation is a consequence of the special 
form of the Hamiltonian governing the D-particle 
dynamics. The nature of the space-time uncertainty 
relation is thus similar to that of the ordinary time-energy 
uncertainty relation, reflecting the first derivation 
of the former as a reinterpretation of the 
latter. 
On the other hand, in the case of the D-instantons, 
the time direction  is treated as a matrix variable 
which is of equal footing as the spatial directions. 
The space-time uncertainty relation is taken into account 
by imposing the $\delta$-function constraint as a 
quantum condition. 
 I expect that something which unifies both approaches 
and has some higher symmetry structure than the $SU(N)$ of the 
effective actions in order to allow us the light-cone 
gauge fixing  is 
necessary for constructing the covariant (11 dimensional) 
formulation of the M-theory in terms of a matrix model. 
It seems that this requires us to invent 
a totally new way of formulating dynamics, as is already 
expressed earlier in this talk. 

To conclude, I have discussed  the relevance and 
possible formulations of a simple space-time uncertainty principle. 
Although I think the discussions provide some sound evidence,  
there remains much to be further clarified. 
Also it is very important to try to combine our considerations with 
other  proposals which are 
related to black-hole physics in string theory, 
 like e.g., `black hole complementarity' \cite{susskind2}, 
`holographic principle' \cite{hol}, `correspondence principle for 
black holes and strings' \cite{horopol}. The black hole is  
expected to be 
one of the most fruitful testing grounds for the 
nonperturbative string theory.    

\section{Acknowledgements}
I would like to thank the organizers 
for inviting me to participate in this interesting 
conference. 
The present work is partially supported by 
Grant-in-Aid for Scientific  Research (No. 09640337) 
from the Ministry of  Education, Science and Culture.

\eject

\end{document}